\documentclass[
]{ceurart}
\usepackage{listings}
\usepackage{changepage}
\usepackage{tikz}
\usepackage{pgfplots}
\usepackage{float}
\usetikzlibrary{patterns}
\usepackage[breakable, skins]{tcolorbox}
\newenvironment{summarybox}
{\begin{tcolorbox}
[breakable,enhanced,arc=0mm,colback=gray!10,frame hidden,overlay broken={%
    \draw[thick,black] (interior.north west)--(interior.south west);
},overlay unbroken={%
    \draw[thick,black] (interior.north west)--(interior.south west);
},left=2pt,right=0pt,top=0pt,bottom=0pt,before={\vspace{3pt}\noindent},after={\vspace{0pt}}]
\setlength{\baselineskip}{0.75\baselineskip}}
{\end{tcolorbox}}
\newenvironment{summary}
{\vspace{5pt}\noindent\begin{summarybox}}
{\end{summarybox}\vspace{-5pt}}

\pgfplotsset{compat=1.18}
\begin{document}

\copyrightyear{2026}
\copyrightclause{Copyright for this paper by its authors. Use permitted under Creative Commons License Attribution 4.0 International (CC BY 4.0).}

\conference{BENEVOL2026: Belgium-Netherlands Software Evolution Workshop
26--27 November, 2026, Madrid, Spain.}

\title{Does Confusing Code Confuse LLMs? A Pilot Study on C++}

\author[1]{Tim van de Wetering}[%
%orcid=0000-0000-0000-0000,
%email=t.vandewetering@student.utwente.nl,
%url=https://yamadharma.github.io/,
]
\address[1]{Formal Methods \& Tools, University of Twente, Enschede, the Netherlands}
%\address[2]{Computer Science, University of Twente, Enschede, the Netherlands}

\author[1]{Lola Solovyeva}[%
%orcid=0000-0000-0000-0000,
email=o.solovyeva@utwente.nl,
%url=https://kmitd.github.io/ilaria/,
]

\author[1]{Fernando Castor}[%
%orcid=0000-0000-0000-0000,
email=f.castor@utwente.nl,
%url=http://conceptbase.sourceforge.net/mjf/,
]

\begin{abstract}
  Large language models (LLMs) are increasingly deployed for software maintenance tasks that depend heavily on code comprehension. While atoms of confusion,  minimal code patterns empirically shown to impair human program comprehension, are well studied, it remains unclear whether these same patterns affect LLMs, and whether their impact varies across maintenance tasks. We present a pilot study evaluating the impact of atoms of confusion on LLM performance on C++ code drawn from three production libraries, using ChatGPT and GitHub Copilot on three software development tasks: atom injection and removal, output prediction, and code explanation. Our results reveal three preliminary findings: (i) the impact of atoms on LLMs is strongly task-dependent; (ii) atoms requiring reasoning about surrounding code are the most resistant to transformation, while localized syntactic atoms are handled reliably; and (iii) LLMs remove atoms of confusion more reliably than they inject them, suggesting that removal succeeds through pattern matching while injection requires deeper comprehension.

\end{abstract}

% \begin{keywords}
%   LLMs \sep
%   Code Comprehension \sep
%   Atoms of Confusion \sep
%   C++ \sep
%   Software Maintenance
% \end{keywords}

\maketitle

\section{Introduction}
Maintaining existing code accounts for 50–90\% of total software costs~\cite{foster_cost_1993,mynyk_information_2012}, and code comprehension is the bottleneck within that work: developers spend on average 70\% of their time understanding code before they can fix bugs, refactor, or add features~\cite{minelli_know_2015}. Readability is a primary lever on this cost. Clear, well-written code reduces cognitive load and enables efficient maintenance~\cite{weigend_constructing_2010,nguyen_optimizing_2024}, while poor readability, which can be unclear naming, inconsistent formatting, or structural choices, increases comprehension time, error rates during modification, and difficulty finding bugs~\cite{johnson_empirical_2019}. In short, anything that makes code harder to read makes it harder to maintain, and at the scale of the global software industry, small readability effects translate into billions of dollars annually spent on comprehension alone~\cite{foster_cost_1993}.

Large language models (LLMs) are increasingly adopted for code-related development tasks~\cite{schmid_software_2025,bistarelli_usage_2025}, offering significant potential to reduce the cost of software maintenance by assisting with comprehension-dependent tasks such as test generation, bug fixing, refactoring, and documentation. Tools like GitHub Copilot and ChatGPT are now used by millions of developers. However, this rapid adoption comes with challenges around output reliability~\cite{nejjar_llms_2025}: LLMs can generate code that appears correct but contains subtle bugs, security vulnerabilities, or logical errors. Since these models are trained on existing codebases that vary widely in quality and readability, it raises the question if the same readability factors that affect humans also affect LLM performance. Understanding this relationship is essential for deploying these tools effectively in practice.

While the effect of readability on human code comprehension is well established~\cite{gopstein_thinking_2020, gopstein_understanding_2017, gopstein_prevalence_2018, medeiros_investigation_2019, tahsin_prevalence_2023, torres_investigation_2023}, its effect on LLMs appears different: LLMs do not benefit from human-oriented formatting~\cite{pan_hidden_2025}, yet factors such as structure, complexity, and clarity do improve their performance~\cite{ouedraogo_humanaligned_2025}. No study has systematically evaluated whether confusing code patterns known to affect humans also affect LLMs. Such patterns are known as \textit{atoms of confusion}, which are minimal code constructs empirically validated to cause programmers to misunderstand source code, each with a functionally equivalent, clearer alternative~\cite{gopstein_understanding_2017}. Multiple studies have shown they substantially increase human error rates and comprehension time~\cite{gopstein_prevalence_2018, torres_investigation_2023}, with some patterns raising error rates by as much as 60\%~\cite{gopstein_understanding_2017}. Since LLMs are trained on codebases containing these patterns, whether they impair LLM comprehension in the same way is an open and practically important question.

%Not happy with it, might change later
This study investigates how atoms of confusion impact LLM performance on practical software development tasks. Understanding this will allow for more effective usage of LLMs in development workflows and guide improvements to make models more robust. This study focuses on C++, a language that LLMs have more trouble with compared to more readable languages like Python. Using a language that LLMs already struggle with provides a more useful testing environment, as languages LLMs perform well on could reduce the observable impact of atoms of confusion.

\section{Background \& Related work}
\par Gopstein at al.~\cite{gopstein_understanding_2017} introduced atoms of confusion, minimal code patterns that reliably cause programmers to misunderstand source, through empirical experiments with human programmers. Their approach involved extracting candidate patterns from known confusing code, such as winners of the International Obfuscated C Code Contest (IOCCC), and validating these candidates through controlled experiments where programmers hand-evaluate small code snippets. Confusion is operationally defined as occurring when a programmer's interpretation of code behaviour differs from the actual behaviour specified by the language. For every atom there exists a functionally equivalent alternative that is significantly less confusing. Following study of Gopstein et al.~\cite{gopstein_prevalence_2018} demonstrated that atoms occur frequently in production software, analyzing 14 major C/C++ projects, including the Linux kernel and GCC~\cite{gopstein_prevalence_2018}. They found atoms appearing once every 23 lines on average, totaling over 3.6 million occurrences. Bug-fixing commits were 1.25 times more likely to remove atoms than other commits, which suggests that there is a relationship between the usage of these patterns and the occurrence of (certain) bugs in software. Additionally, atoms were 1.13 times more likely to be commented than other code, indicating that developers recognize their potential for confusion.
\par Moon et al.~\cite{moon_dont_2025} showed that LLMs exhibit systematic biases when evaluating code correctness based on superficial variations such as authorship claims, self-declared correctness, and variable naming, with accuracy dropping by up to 26.7 percentage points under biased conditions. Their work focuses on surface-level features that do not affect code semantics. Instead, we examine structural patterns, atoms of confusion, that affect readability while preserving functional equivalence. Pan et al.~\cite{pan_hidden_2025} conducted a large-scale study of whether formatting elements such as indentation and newlines affect LLM performance on code completion, across four languages and ten LLMs~\cite{pan_hidden_2025}. They concluded that LLMs do not benefit from human-readability formatting in the same way humans do. Their work examines visual formatting elements that do not alter semantics. However, we examine structural patterns that alter readability and may alter how the code is understood. Ouédraogo et al.~\cite{ouedraogo_humanaligned_2025} introduced a large-scale benchmark for LLM-based code readability assessment (1.4M evaluations across 10 LLMs and 3 languages), finding that developer-informed prompts improve aspect coverage but increase score variability. Their work evaluates LLMs as readability assessors, where we investigate whether poor readability itself degrades LLM performance on comprehension-dependent tasks, regardless of whether readability is explicitly considered. Abdelsalam et al.~\cite{abdelsalam_how_2025} compared how LLMs and humans process confusing code, finding that LLM perplexity spikes correlate spatially and in amplitude with human neurophysiological responses to atoms of confusion. Their work shows that LLMs exhibit confusion patterns similar to humans. So, complementary we investigate whether these patterns actually impair LLM performance on downstream software development tasks.

\begin{table}[t]
    \begin{adjustwidth}{-1cm}{-1cm}
        \centering
        \caption{
        Selected Atoms of Confusion from original study of Gopstein et al.~\cite{gopstein_understanding_2017}}
        \label{tab:atoms_of_confusion}
        \footnotesize
        \begin{tabular}{lp{5.5cm}p{3cm}p{3cm}}
        \hline
        \textbf{Atom} & \textbf{Description} & \textbf{Example} & \textbf{Transformed} \\
        \hline
        Pre-/Post-Increment & Post-increment returns the original value before incrementing. Pre-increment first updates then returns the new value. & \verb|V1 = V2++;| & \verb|V1 = V2; V2+=1;| \\
        \hline
        Operator Precedence & Most programmers know only a subset of C's precedence rules. & \verb|0 && 1 || 2| & \verb|(0 && 1) || 2| \\
        \hline
        Conditional Operator & The ternary operator functions like if/else but as an expression. & \verb|V2=(V1==3)?2:V2| & \verb|if(V1==3) V2=2;| \\
        \hline
        Type Conversion & Implicit type conversion may produce unexpected outcomes. & \verb|(double)(3/2)| & \verb|trunc(3.0/2.0)| \\
        \hline
        Omitted Curly Braces & Without braces, the scope of control statements is unclear. & \verb|if(V) F(); G();| & \verb|if(V){F();}G();| \\
        % \hline
        % Logic as Control Flow & Short-circuiting allows \&\& and || to be used for conditional execution. & \verb|V1 && F2();| & \verb|if (V1) F2();| \\
        % \hline
        % Arithmetic as Logic & Arithmetic operators can mimic logical predicates but imply non-boolean values. & \verb|(V1-3)*(V2-4)| & \verb|V1!=3 && V2!=4| \\
        \hline
        Repurposed Variables & Using a variable in different roles makes its meaning difficult to follow. & \verb|argc = 7;| & \verb|int V1 = 7;| \\
        \hline
        Change of Literal Encoding & Different numeric representations affect understandability. & \verb|printf("%d",013)| & \verb|printf("%d",11)| \\
        \hline
        Comma Operator & The comma operator sequences computations but is often misinterpreted. & \verb|V3=(V1+=1, V1)| & \verb|V1+=1; V3=V1;| \\
        \hline
        Assignment as Value & Assignment returns a value in addition to changing state. & \verb|V1 = V2 = 3;| & \verb|V2=3; V1=V2;| \\
        \hline
        \end{tabular}
    \end{adjustwidth}
\end{table}

\section{Methodology}
\par This study investigates how atoms of confusion, code patterns empirically validated to hinder human program comprehension, affect LLM performance across different software development tasks. Therefore, we focus on the following research question:
\begin{quote}
\textit{To what extent do 
atoms of confusion also affect LLM's ability to understand code?}
\end{quote}
To answer this question, we evaluate LLM performance on software development tasks using code snippets that contain atoms of confusion from the original study~\cite{gopstein_understanding_2017}, and compare it to their performance on functionally equivalent snippets in which the atoms have been removed.
\par \textbf{\textit{Software development tasks.}} Three tasks were chosen to span different levels of reasoning depth and to reflect common ways LLMs are used in maintenance workflows. This allows us to evaluate whether the impact varies with the nature of the task.
\begin{itemize}
    \item \textbf{\textit{Atom injection and removal:}}  the LLMs are prompted to introduce or eliminate a specific atom of confusion in a given code snippet while preserving its behavior. This task probes whether the model can manipulate confusing patterns as structural objects, which is a prerequisite for LLM-assisted refactoring and readability improvements.
    \item \textbf{\textit{Output prediction:}} the LLMs are prompted to predict the output of a program given a specific input. This task requires the model to simulate program execution, and is therefore sensitive to comprehension errors. It reflects tasks such as debugging, test generation, and etc.
    \item \textbf{\textit{Code explanation:}}  the LLMs are prompted to describe a code snippet's logic, control flow, and behavior in natural language. This task reflects a common use of LLMs in practice: helping developers understand unfamiliar code. Comparing performance on this task against output prediction lets us assess whether atom impact varies with the reasoning depth a task demands.
\end{itemize}
\par \textbf{\textit{Selected atoms of confusion.}} From the set of candidates for atoms identified in multiple studies~\cite{gopstein_thinking_2020, gopstein_understanding_2017, gopstein_prevalence_2018, medeiros_investigation_2019, tahsin_prevalence_2023, torres_investigation_2023}, the most impactful instances, the instances with the largest effect size, and the instances that occur most frequently were selected and presented in Table~\ref{tab:atoms_of_confusion}.  
%In addition to these established atoms, we explored several C++ specific atom candidates to determine whether certain language-specific patterns exhibit similar confusion properties.
We also include one C++ specific candidate, 
%as it represents a minimal confusing pattern with a functionally equivalent, clearer alternative. 
\textbf{SFINAE vs Concepts}. 
%is a C++ specific candidate we add to the set of established atoms. 
SFINAE, Substitution Failure Is Not An Error, typically used through the \texttt{enable\_if} idiom, conditionally enables template functions via cryptic syntax that hijacks return types or adds dummy parameters. C++20 introduced concepts explicitly to replace it, after more than two decades of community complaints about unreadable code and opaque error messages~\cite{voutilainen2016concepts,smith2017conceptsts}. Stroustrup himself described \texttt{enable\_if}-based overloading as "error-prone and tedious assembly programming"~\cite{stroustrup2017goodconcepts}. %providing strong evidence that this pattern meets the definition of an atom of confusion.
\par \textbf{\textit{Dataset construction.}} We construct paired code samples that differ only in the presence or absence of a single atom of confusion while remaining functionally equivalent. Samples are drawn from three widely used, well-tested C++ libraries: \href{https://github.com/ocornut/imgui}{ImGui}, a UI library used across the game and software industry; \href{https://github.com/boostorg/qvm}{Boost.QVM} , a header-only linear algebra library from the Boost ecosystem; and \href{https://github.com/ericniebler/range-v3}{ranges-v3}, the template-heavy reference implementation of C++20 ranges, chosen specifically because it exercises advanced idioms such as SFINAE. If the sample does not already have an atom of confusion, then they are introduced through code transformation with scripts and in some cases using LLMs from this study. All transformed samples are validated through compilation and each library's existing test suite to ensure syntactic correctness and functional equivalence with the original. Transformations that fail these checks are excluded. 
%This validation ensures that any performance differences we observe reflect the presence of the atom rather than unintended behavioral changes.
\par \textbf{\textit{Models \& Experimental setup}}. We evaluate two freely accessible frontier LLMs: ChatGPT (GPT-5.3) and GitHub Copilot (Claude Haiku 4.5). Both are the default free-tier models for their respective platforms.
%reflecting realistic usage for a broad audience. 
For output prediction we additionally evaluate Gemini Flash 3.5, since the smaller sample size for that task made a third model feasible. For atom injection and removal, LLMs were prompted to transform 20 samples per atom. The transformed samples replaced the original in the source library and were validated using the library's existing test suite. For output prediction, we used 5 samples per atom, modified where necessary to expose a return value, and prompted the LLMs in a zero-shot condition. For the code explanation task, the same 20 samples per atom used in the transformation task were provided to the LLMs. The models were instructed to explain each code snippet in terms of its overall purpose, control flow, key computations, and output behavior. 
\par \textbf{\textit{Metrics}}. For atom injection and removal we use two complementary metrics. Accuracy measures the percentage of transformed samples that still pass the library's test suite, capturing whether the transformation preserved behavior. Transformation Success Rate (TSR) measures the percentage of atoms correctly transformed, capturing whether the intended change was actually performed. The two metrics can diverge: a transformation that leaves the code unchanged achieves 100\% accuracy but 0\% TSR, while a partial or incorrect transformation may still pass the tests. We report both because each answers a different question. For output prediction, we report the number of correct predictions per atom out of the total samples. For code explanation, explanations are manually graded per category as correct, incorrect (factual errors), or neither (not enough information for evaluation).

\section{Results \& Discussion}

\begin{figure}
    \centering
    \includegraphics[width=\linewidth]{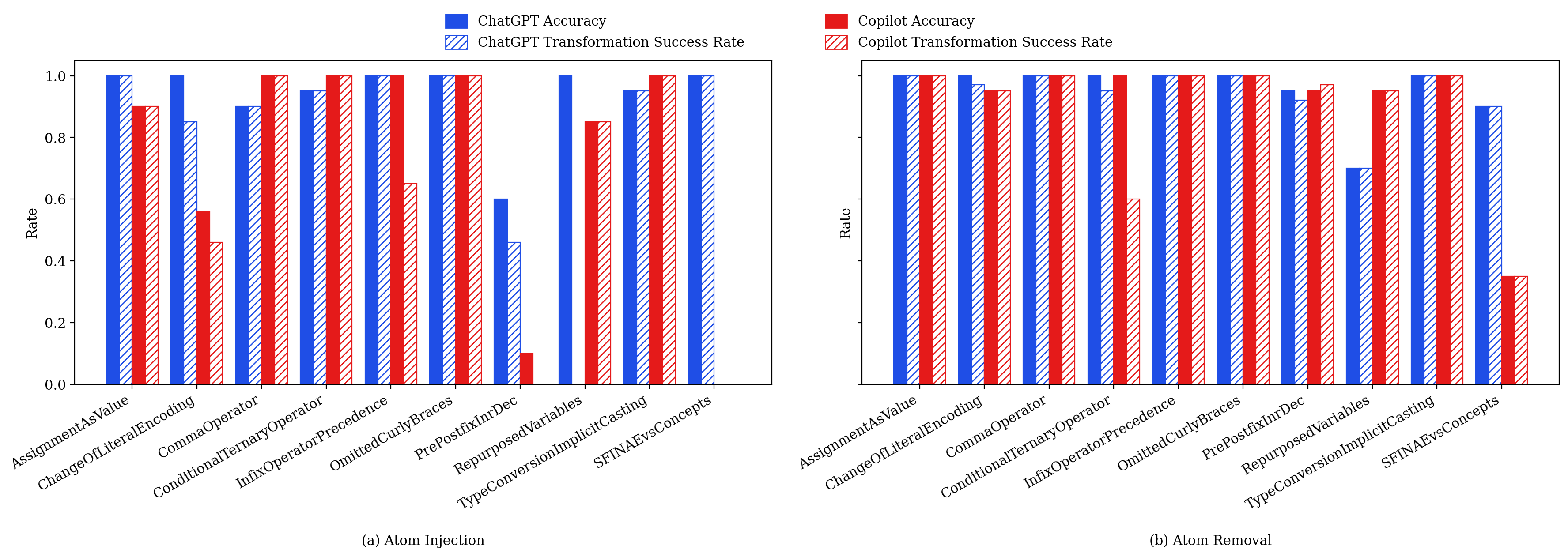}
    \caption{
    \footnotesize Results of injecting and removing atoms of confusion in code snippets using ChatGPT and Copilot, shown in blue and red, respectively. Solid bars indicate accuracy, measured as the proportion of transformed code snippets that pass the tests, while dashed bars indicate the transformation success rate, i.e., whether the atom of confusion was successfully injected or removed.
}
    \label{fig:injection_removal}
\end{figure}

\par \textbf{\textit{LLMs remove atoms of confusion more reliably than they inject them.}}  Figure~\ref{fig:injection_removal} presents that removing atoms is generally easier for LLMs than injecting them. Focusing on the TSR, the models particularly struggle to inject \textit{ChangeOfLiteralEncoding}, \textit{InfixOperatorPrecedence}, \textit{PrePostfixIncDec}, \textit{RepurposedVariables}, and \textit{SFINAEvsConcepts}. 
%We use TSR as the primary measure of transformation effectiveness because it captures whether the model successfully performed the intended transformation, and whether the target atom of confusion was actually injected or removed. In contrast, accuracy reflects whether the transformation preserved the remaining behavior of the code snippet, as determined by whether the transformed code continues to pass the test suite. 
The distinction between the TSR and accuracy shows for \textit{RepurposedVariables} injection, where ChatGPT achieves 100\% accuracy but 0\% TSR. The model returned the original code unchanged in all samples, effectively refusing the transformation while still passing tests. A different pattern emerges for atom removal: the models successfully remove most atoms of confusion, with the main exceptions being \textit{ConditionalTernaryOperator}, \textit{RepurposedVariables}, and \textit{SFINAEvsConcepts}. %Taken together, these results reveal an asymmetry between the two transformation directions. 
One plausible explanation is that removing an atom often only requires recognizing a known pattern and replacing it with an equivalent, while injecting an atom requires deeper comprehension via understanding the surrounding code well enough to introduce the construct at a semantically valid location without altering behavior. 
\par \textit{\textbf{Atoms requiring reasoning about surrounding code are the most resistant to transformation.}} The atoms most resistant to injection share a common property: they require reasoning about surrounding code rather than local rewriting. \textit{SFINAEvsConcepts} requires understanding template constraints, \textit{RepurposedVariables} requires understanding how a variable's role changes across a function, \textit{Pre-/Postfix Increment/Decrement} requires reasoning about evaluation order. In contrast, atoms that are handled reliably in both directions — \textit{OmittedCurlyBraces}, \textit{AssignmentAsValue}, \textit{CommaOperator} — involve localized syntactic transformations. This suggests that LLMs can perform mechanical rewrites reliably but struggle with transformations that demand understanding of surrounding context.
\begin{figure}
    \centering
    \includegraphics[width=0.85\linewidth]{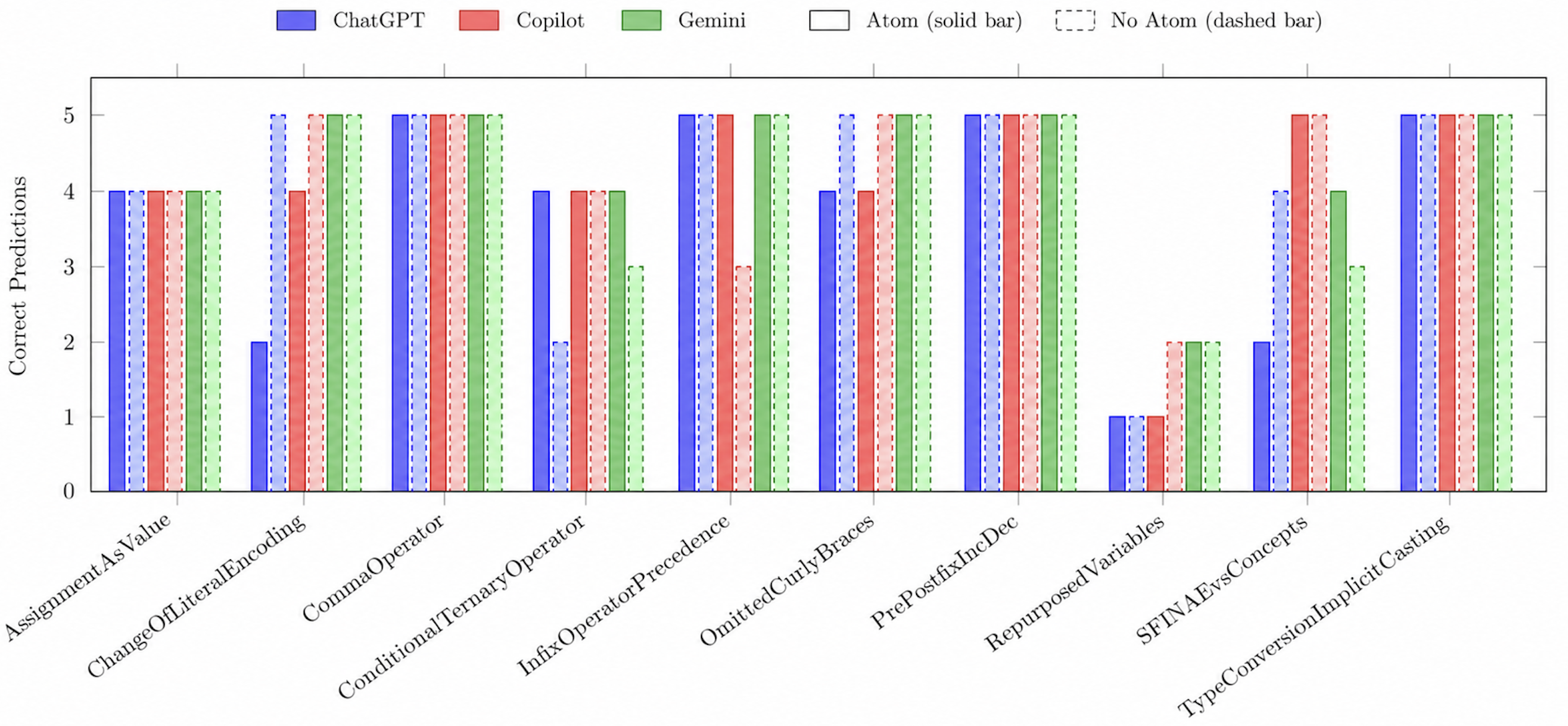}
    \caption{
    \footnotesize Results of the output prediction task for ChatGPT, Copilot, and Gemini. Bar color indicates the model selected, while line style indicates whether the code snippet contained an atom of confusion: solid bars represent snippets containing an atom of confusion, and dashed bars represent snippets without one.
 }
    \label{fig:output_prediction}
\end{figure}
\par \textit{\textbf{The impact of atoms of confusion on LLMs is task-dependent}}. Figure~\ref{fig:output_prediction} compares model performance on code snippets with and without atoms of confusion, allowing us to assess their impact on the ability of models to correctly predict program output. The most notable performance degradation is observed for \textit{ChangeOfLiteralEncoding}, where ChatGPT score decreases from 5 to 2 and Copilot from 5 to 4. For \textit{OmittedCurlyBraces}, the scores of both ChatGPT and Copilot decrease from 5 to 4. A substantial decline is also observed for \textit{SFINAEvsConcepts}, for which ChatGPT score drops from 4 to 2. Several atoms produced no observable degradation for any model, including \textit{CommaOperator}, \textit{InfixOperatorPrecedence}, and \textit{Pre-/Postfix Increment/Decrement}, despite being known sources of confusion for human programmers. \textit{RepurposedVariables} is excluded from this discussion, as performance is uniformly low in both cases, indicating that the atom is not the source of difficulty in these samples.
\par In contrast, the results for the code explanation task show a different pattern, with accuracy remaining at 100\% across nearly all evaluated atoms of confusion and models. Given the consistently high and nearly identical performance across all models, we omit the corresponding plot. The models were generally able to identify the purpose, control flow, and behaviour of the code correctly even when confusing patterns were present. The only exception is \textit{ChangeOfLiteralEncoding}, for which the accuracy decreases to 90\% for code snippets containing the atom, the same atom that produced the largest output-prediction drop, and in both cases the errors stemmed from misreading octal literals as decimal values. 
\par These findings suggest that tasks requiring natural-language explanations are considerably more robust to atoms of confusion than tasks requiring precise reasoning about program execution. One possible explanation is that code explanation closely resembles the types of tasks on which modern LLMs are extensively trained. Large amounts of source code, documentation, comments, and question-answer pairs expose models to many examples of describing code behavior in natural language, allowing them to recognize common programming patterns without necessarily performing detailed execution reasoning. Therefore, generating high-level explanations may rely more on retrieving and generalizing learned representations than on accurately simulating program execution. 
\begin{summary}
\footnotesize
    \par \textbf{\textit{Implications.}} Our results show that the models reliably remove confusing code patterns. Hence, they could help clean up, refactor, and simplify code, for example in legacy codebases. Patterns that depend on the surrounding code, such as reused variables or the move from SFINAE to C++20 concepts, still need a developer to review the result. This matters for maintenance because such changes often come up when modernizing older codebases.
\end{summary}

%%
%% Define the bibliography file to be used
\small
\bibliography{main}

@misc{abdelsalam_how_2025,
  title = {How Do {{Humans}} and {{LLMs Process Confusing Code}}?},
  author = {Abdelsalam, Youssef and Peitek, Norman and Maurer, Anna-Maria and Toneva, Mariya and Apel, Sven},
  year = 2025,
  month = aug,
  number = {arXiv:2508.18547},
  eprint = {2508.18547},
  primaryclass = {cs},
  publisher = {arXiv},
  doi = {10.48550/arXiv.2508.18547},
  archiveprefix = {arXiv}
}

@article{bistarelli_usage_2025,
  title = {Usage of {{Large Language Model}} for {{Code Generation Tasks}}: {{A Review}}},
  shorttitle = {Usage of {{Large Language Model}} for {{Code Generation Tasks}}},
  author = {Bistarelli, Stefano and Fiore, Marco and Mercanti, Ivan and Mongiello, Marina},
  year = 2025,
  month = jul,
  journal = {SN Computer Science},
  volume = {6},
  number = {6},
  pages = {673},
  issn = {2661-8907},
  doi = {10.1007/s42979-025-04241-5},
  urldate = {2025-12-15},
  langid = {english}
}

@phdthesis{foster_cost_1993,
  type = {Doctoral},
  title = {Cost Factors in Software Maintenance.},
  author = {Foster, John R.},
  year = 1993,
  url = {https://etheses.dur.ac.uk/1561/},
  urldate = {2025-12-09},
  school = {Durham University}
}

@inproceedings{gopstein_prevalence_2018,
  title = {Prevalence of Confusing Code in Software Projects: Atoms of Confusion in the Wild},
  shorttitle = {Prevalence of Confusing Code in Software Projects},
  booktitle = {Proceedings of the 15th MSR},
  author = {Gopstein, Dan and Zhou, Hongwei Henry and Frankl, Phyllis and Cappos, Justin},
  year = 2018,
  month = may,
  series = {{{MSR}} '18},
  doi = {10.1145/3196398.3196432},
  isbn = {978-1-4503-5716-6}
}

@inproceedings{gopstein_thinking_2020,
  title = {Thinking Aloud about Confusing Code: A Qualitative Investigation of Program Comprehension and Atoms of Confusion},
  shorttitle = {Thinking Aloud about Confusing Code},
  booktitle = {Proceedings of the 28th {ESEC/FSE}},
  author = {Gopstein, Dan and others},
  year = 2020,
  month = nov,
}

@inproceedings{gopstein_understanding_2017,
  title = {Understanding Misunderstandings in Source Code},
  booktitle = {Proceedings of the 2017 11th FSE},
  author = {Gopstein, Dan and Iannacone, Jake and Yan, Yu and DeLong, Lois and Zhuang, Yanyan and Yeh, Martin K.-C. and Cappos, Justin},
  year = 2017,
  month = aug,
  series = {{{ESEC}}/{{FSE}} 2017},
  doi = {10.1145/3106237.3106264},
  isbn = {978-1-4503-5105-8}
}

@inproceedings{johnson_empirical_2019,
  title = {An {{Empirical Study Assessing Source Code Readability}} in {{Comprehension}}},
  booktitle = {2019 IEEE ICSME},
  author = {Johnson, John and Lubo, Sergio and Yedla, Nishitha and Aponte, Jairo and Sharif, Bonita},
  year = 2019,
  month = sep,
  pages = {513--523},
  issn = {2576-3148},
  doi = {10.1109/ICSME.2019.00085},
  urldate = {2025-11-01}
}

@article{medeiros_investigation_2019,
  title = {An Investigation of Misunderstanding Code Patterns in {{C}} Open-Source Software Projects},
  author = {Medeiros, Fl{\'a}vio and Lima, Gabriel and Amaral, Guilherme and Apel, Sven and K{\"a}stner, Christian and Ribeiro, M{\'a}rcio and Gheyi, Rohit},
  year = 2019,
  month = aug,
  journal = {Empirical Software Engineering},
  volume = {24},
  number = {4},
  pages = {1693--1726},
  issn = {1573-7616},
  doi = {10.1007/s10664-018-9666-x},
  langid = {english}
}

@inproceedings{minelli_know_2015,
  title = {I {{Know What You Did Last Summer}} - {{An Investigation}} of {{How Developers Spend Their Time}}},
  booktitle = {2015 {{IEEE}} 23rd ICPC},
  author = {Minelli, Roberto and Mocci, Andrea and Lanza, Michele},
  year = 2015,
  month = may,
  pages = {25--35},
  issn = {1092-8138},
  doi = {10.1109/ICPC.2015.12},
  urldate = {2025-11-01}
}

@misc{moon_dont_2025,
  title = {Don't {{Judge Code}} by {{Its Cover}}: {{Exploring Biases}} in {{LLM Judges}} for {{Code Evaluation}}},
  shorttitle = {Don't {{Judge Code}} by {{Its Cover}}},
  author = {Moon, Jiwon and Hwang, Yerin and Lee, Dongryeol and Kang, Taegwan and Kim, Yongil and Jung, Kyomin},
  year = 2025,
  month = may,
  doi = {10.48550/arXiv.2505.16222},
  archiveprefix = {arXiv}
}

@phdthesis{mynyk_information_2012,
  title = {Information {{Technology Programming Standards}} and {{Annual Project Maintenance Costs}}},
  author = {Mynyk, John},
  year = 2012,
  journal = {ProQuest LLC},
  urldate = {2025-11-10},
  isbn = {978-1-303-72316-2},
  langid = {english},
  school = {University of Phoenix}
}

@article{nejjar_llms_2025,
  title = {{{LLMs}} for Science: {{Usage}} for Code Generation and Data Analysis},
  shorttitle = {{{LLMs}} for Science},
  author = {Nejjar, Mohamed and Zacharias, Luca and Stiehle, Fabian and Weber, Ingo},
  year = 2025,
  journal = {Journal of Software: Evolution and Process},
  volume = {37},
  number = {1},
  pages = {e2723},
  issn = {2047-7481},
  doi = {10.1002/smr.2723},
  urldate = {2025-11-01},
  copyright = {\copyright{} 2024 The Author(s). Journal of Software: Evolution and Process published by John Wiley \& Sons Ltd.},
  langid = {english}
}

@misc{nguyen_optimizing_2024,
  type = {fi={{AMK-opinn\"aytety\"o}}\textbar sv={{YH-examensarbete}}\textbar en={{Bachelor}}'s Thesis\textbar},
  title = {Optimizing {{Developer Experience Through Clean Code}}},
  author = {Nguyen, Tai Phat},
  year = 2024,
  url = {http://www.theseus.fi/handle/10024/879121},
  urldate = {2025-12-09},
  copyright = {fi=All rights reserved. This publication is copyrighted. You may download, display and print it for Your own personal use. Commercial use is prohibited.\textbar sv=All rights reserved. This publication is copyrighted. You may download, display and print it for Your own personal use. Commercial use is prohibited.\textbar en=All rights reserved. This publication is copyrighted. You may download, display and print it for Your own personal use. Commercial use is prohibited.\textbar},
  langid = {english}
}

@misc{ouedraogo_humanaligned_2025,
  title = {Human-{{Aligned Code Readability Assessment}} with {{Large Language Models}}},
  author = {Ou{\'e}draogo, Wendk{\^u}uni C. and Li, Yinghua and Dang, Xueqi and Borsukiewicz, Pawel and Zhou, Xin and Koyuncu, Anil and Klein, Jacques and Lo, David and Bissyand{\'e}, Tegawend{\'e} F.},
  year = 2025,
  doi = {10.48550/arXiv.2510.16579},
  archiveprefix = {arXiv}
}

@misc{pan_hidden_2025,
  title = {The {{Hidden Cost}} of {{Readability}}: {{How Code Formatting Silently Consumes Your LLM Budget}}},
  shorttitle = {The {{Hidden Cost}} of {{Readability}}},
  author = {Pan, Dangfeng and Sun, Zhensu and Zhang, Cenyuan and Lo, David and Du, Xiaoning},
  year = 2025,
  month = aug,
  doi = {10.48550/arXiv.2508.13666},
  archiveprefix = {arXiv}
}

@misc{schmid_software_2025,
  title = {Software {{Architecture Meets LLMs}}: {{A Systematic Literature Review}}},
  shorttitle = {Software {{Architecture Meets LLMs}}},
  author = {Schmid, Larissa and Hey, Tobias and Armbruster, Martin and Corallo, Sophie and Fuch{\ss}, Dominik and Keim, Jan and Liu, Haoyu and Koziolek, Anne},
  year = 2025,
  month = may,
  eprint = {2505.16697},
  primaryclass = {cs},
  doi = {10.48550/arXiv.2505.16697},
  urldate = {2025-12-15}
}

@misc{smith2017conceptsts,
  title = {{{P0726R0}}: {{Does}} the {{Concepts TS Improve}} on {{C}}++17?},
  author = {{Duncan P. N. Exon Smith}},
  url = {https://www.open-std.org/jtc1/sc22/wg21/docs/papers/2017/p0726r0.html},
  urldate = {2025-10-31}
}

@article{stroustrup2017goodconcepts,
  title = {How to Design Good Concepts and Use Them Well},
  author = {Stroustrup, Bjarne},
  year = 2017,
  month = jan,
  url = {https://www.stroustrup.com/good_concepts.pdf},
  langid = {english}
}

@article{tahsin_prevalence_2023,
  title = {Prevalence of `{{Atoms}} of {{Confusion}}' in {{Open Source Java Systems}}: {{An Empirical Study}}},
  shorttitle = {Prevalence of `{{Atoms}} of {{Confusion}}' in {{Open Source Java Systems}}},
  author = {Tahsin, Noshin and Fuad, Nafis and Satter, Abdus},
  year = 2023,
  month = feb,
  url = {https://www.authorea.com/doi/full/10.22541/au.167570695.54470176?commit=d9faaece30c7d2ce7abf98ae8a723053ea51c058},
  urldate = {2025-10-31}
}

@article{torres_investigation_2023,
  title = {An {{Investigation}} of Confusing Code Patterns in {{JavaScript}}},
  author = {Torres, Adriano and others},
  year = 2023,
  month = sep,
  journal = {JSS},
  volume = {203},
  pages = {111731},
  issn = {0164-1212},
}

@misc{voutilainen2016concepts,
  title = {Why {{I}} Want {{Concepts}}, but Why They Should Come Later Rather than Sooner},
  author = {Calabrese, Matt},
  url = {https://www.open-std.org/jtc1/sc22/wg21/docs/papers/2016/p0240r0.html},
  urldate = {2025-10-31}
}

@article{weigend_constructing_2010,
  title = {Constructing {{Complex Things}} without {{Getting Confused}} - {{Programming Techniques}} and {{Reduction}} of {{Cognitive Load}}},
  author = {Weigend, Michael},
  year = 2010,
  langid = {english}
}

\end{document}